\documentclass[%
 reprint,
 amsmath,amssymb,
 aps,
]{revtex4-2}
\usepackage{tipa}
\usepackage[utf8]{inputenc}
\usepackage{siunitx}
\usepackage{bm}
\usepackage{xcolor}
\usepackage{graphicx}
\usepackage{dcolumn}
\usepackage{bm}

\begin{document}

\preprint{APS/123-QED}

\title{Spectral splitting and concentration of broadband light using neural networks}

\author{Alim Yolalmaz$^{1, 2, 3}$}

\affiliation{
 $^{1}$Programmable Photonics Group, Department of Physics, Middle East Technical University, 06800 Ankara, Turkey\\
 $^{2}$Micro and Nanotechnology Program, Middle East Technical University, 06800 Ankara, Turkey\\
 $^{3}$The Center for Solar Energy Research and Applications (G\"{U}NAM), Middle East Technical University, 06800 Ankara, Turkey
}%

\author{Emre Y\"{u}ce$^{1, 2, 3,}$}
 \email{eyuce@metu.edu.tr}
\affiliation{
  $^{1}$Programmable Photonics Group, Department of Physics, Middle East Technical University, 06800 Ankara, Turkey\\
 $^{2}$Micro and Nanotechnology Program, Middle East Technical University, 06800 Ankara, Turkey\\
 $^{3}$The Center for Solar Energy Research and Applications (G\"{U}NAM), Middle East Technical University, 06800 Ankara, Turkey
}%

\date{\today}

\begin{abstract}

Compact photonic elements that control both the diffraction and interference of light offer superior performance at ultra-compact dimensions. Unlike conventional optical structures, these diffractive optical elements can provide simultaneous control of spectral and spatial profile of light. However, the inverse-design of such a diffractive optical element is time-consuming with current algorithms, and the designs generally lack experimental validation. Here, we develop a neural network model to experimentally design and validate SpliCons; a special type of diffractive optical element that can achieve spectral splitting and simultaneous concentration of broadband light. We use neural networks to exploit nonlinear operations that result from wavefront reconstruction through a phase plate. Our results show that the neural network model yields enhanced spectral splitting performance for phase plates with quantitative assessment compared to phase plates that are optimized via local search optimization algorithm. The capabilities of the phase plates optimized via neural network are experimentally validated by comparing the intensity distribution at the output plane. Once the neural networks are trained, we manage to design SpliCons with 96.6 $\pm$ 2.3\% accuracy within 2 seconds, which is orders of magnitude faster than iterative search algorithms. We openly share the fast and efficient framework that we develop in order to contribute to the design and implementation of diffractive optical elements that can lead to transformative effects in microscopy, spectroscopy, and solar energy applications.

\end{abstract}

\maketitle


\section{\label{sec:level1}Introduction}

Miniaturized optical elements is an advancing research field aimed to reduce size, weight, and cost of optical systems in the meantime enhancing performance in a variety of application areas such as controlling phase, polarization \cite{Arbabi2015} and absorption \cite{Sell2016}, of light beams in a medium which provide superior performance in spectroscopy \cite{Dana2018}, sensing \cite{Lin1997}, solar energy harvesting \cite{Azad2016}, wavelength demultiplexing \cite{Piggott2015}, particle tracking \cite{Holsteen2019}, imaging \cite{Borhani2018}, image classification \cite{Lin2018}, and quantum computing applications \cite{Kok2007}. One of the promising optical elements is phase plates which, provide control over intensity, polarization, and phase distribution of light with a high degree of freedom \cite{Shutova2019, Kondakci2018, Yang2017, Yolalmaz2020, Guen2021}. Their outperforming functionalities are especially required in spectrally splitting broadband light as conventional lenses lack control in spectral domain \cite{Yolalmaz2020, Yolalmaz2020a, Guen2021}. However, during designing the phase plates, a high number of optimization parameters result in long computation time that seriously hampers their implementation \cite{Jiang2019}.

Spectral and spatial dispersion of broadband light finds diverse application areas as microscopy, digital imaging \cite{Zhou2018}, projection \cite{Jesacher2014}, and solar energy \cite{Widyolar2018, Elikkottil2020}. With the rise in energy demand, intelligent conversion of solar energy is becoming more of a necessity to be addressed fundamentally. Laterally arranged solar cells system has a strong potential in the generation of electricity and incorporate holographic phase plates to achieve spectral splitting of broadband light \cite{Bunthof2018, Stanleya2016}. Unlike conventional diffractive optical elements that are generally designed for one task, SpliCons provide simultaneous spectral splitting and concentration of light \cite{Guen2021}. These multi-functional structures can be optimized with iterative approaches. Still, iterative optimization requires immense computational resources and limits the application of SpliCons due to small numbers of controlled parameters that can yield reduced performance. Instead of iterative approaches, the inverse-design of SpliCons can decrease the optimization time. However, the inverse-design presents several major challenges compared to the iterative: i) phase plates in each frequency of broadband range are needed, ii) combination of these phase plates will still require intermediate phase plates to obtain desired intensity distribution, iii) one-to-many mapping problem \cite{Liu2018}. One-to-many mapping is a big problem due to a data point may be associated with multiple labels instead of a single class \cite{Read2011, Fuernkranz2008}. Thus, spectrally splitting and concentrating the light using the inverse-design is still an unaddressed challenge. Neural network architecture of deep learning could figure out one-to-many mapping problem faced in the inverse-design of SpliCons and provide fast and accurate control over light beams.

Deep learning is a powerful machine learning technique that can perform time-consuming operations using a multilayered neural network within shorter time scales. This technique has shown great success in various optics and photonics tasks such as microscopy \cite{Rivenson2017, Wang2018, Rivenson2018}, imaging \cite{Xue2019, Caramazza2019, Higham2018, Sinha2017}, wavelength demultiplexing \cite{Jiang2019}, metasurface design \cite{Nadell2019, Jiang19}, reconstruction of ultra-short pulses \cite{Zahavy2018}, image classification \cite{Esteva2017}, and beam splitting \cite{Mikhaylov2019}, laser-assisted surface machining \cite{Mills2018, Heath2018}. Moreover, deep learning understands Fourier transform function by using neural neurons having a single layer with a linear transfer function \cite{Velik2008}. 

In this study, we develop for the first time, a neural network model to reconstruct phase patterns for spectrally splitting and spatially concentrating the broadband light and verify our designs experimentally using a spatial light modulator (SLM). In the training procedure of the neural networks, we use a set of known intensity distributions of diffraction patterns and their associated phase plates, where diffraction patterns serve as input and phase plates which structure light are given as output. The results indicate our neural network generates phase patterns for spectrally splitting and spatially concentrating light with high accuracy within a few seconds using a single graphics processing unit (GPU). Our network does not require a manual parameter search to optimize the performance of SpliCons and openly available (see our framework as well as the data set in supplementary) to the community to further accelerate the transformation from uni-functional conventional structures to multi-functional diffraction optical elements.

\section{Methods}
\subsection{Experimental setup}

\begin{figure*}[!htb]
    \centering
    \includegraphics[width=175 mm]{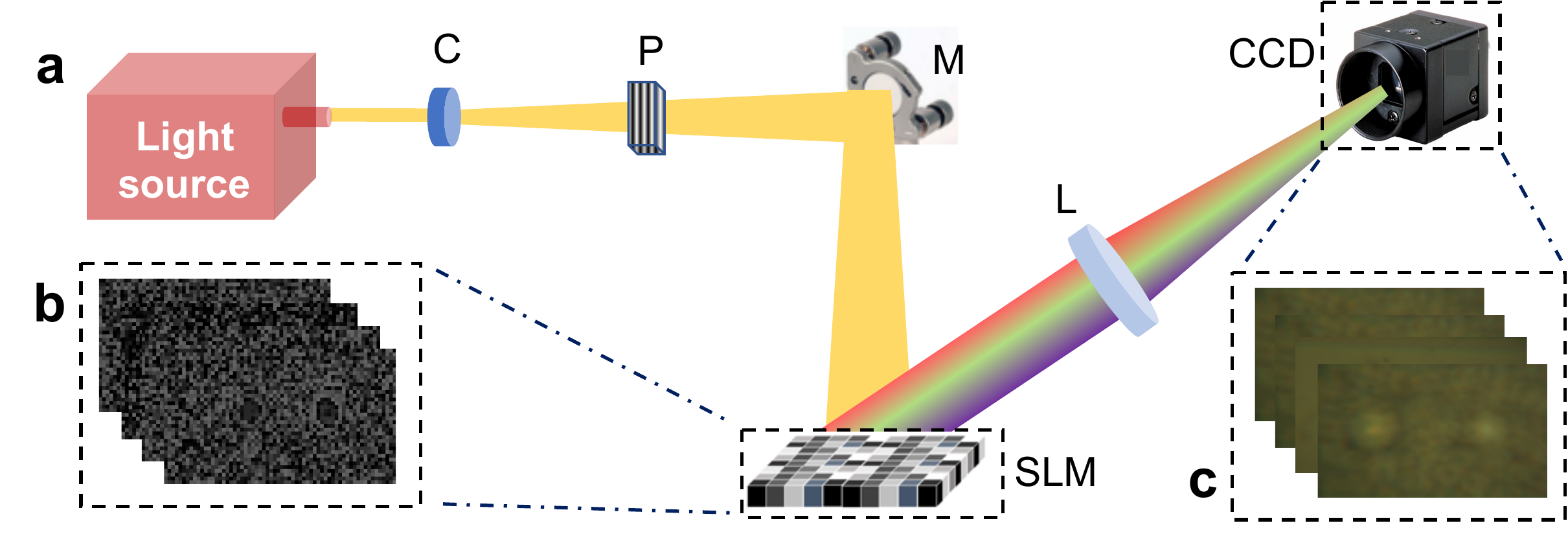}
    \caption{(a) The setup scheme for the spectral splitting and spatially concentrating the broadband light; C; condenser lens, P; linear polarizer, M; mirror, a SLM, L; lens with f= 200 mm, and a CCD camera; (b) Phase patterns that are written on the SLM surface; (c) Intensity distributions measured via the CCD camera.}
    \label{experimental}
\end{figure*}

The setup for spectral splitting and spatially concentrating the broadband light is presented in Fig. \ref{experimental}a. The broadband light source from a Tungsten-Halogen fiber-coupled light source (360 nm – 2600 nm, Optical power: ~7 mW @ 535 nm, Optical power noise: 0.3\%) first passes through an aspheric condenser lens (f: 16 mm). Next, a linear polarizer adjusts the polarization direction of the light so that it is aligned with the SLM modulation axis. Then the light is reflected by a mirror and incident on the SLM (operating between 420 nm - 1100 nm, Holoeye Pluto-NIR-011 phase-only reflective LCOS, frame rate of 60 Hz). The SLM is placed at a small angle to the transmitted light from the mirror and acts as a pixel-wise phase controller object. The SLM that we use here is a phase only SLM and has pixel dimensions \SI{8}{\micro\metre} x \SI{8}{\micro\metre} with a total 1920 x 1080 pixels. Due to long optimization duration of a phase plate to concentrate and spectrally split the broadband light, we grouped pixels of the SLM to a matrix size of 64-by-36 to reduce number of optimized parameters. Each SLM pixel adds at max. 2.28$\pi$ phase shift with 0.23$\pi$ phase steps to the incident light. A phase plate generated by the SLM in the setup controls the phase of the broadband light. The SLM-modulated light passes through a plano-convex lens (f=200 mm) and is collected by the CCD camera (Allied vision, Guppy Pro F-125, the spectral response of the camera chip is given in \cite{Guppy}). The color-CCD camera pixel dimensions are \SI{3.75}{\micro\metre} x \SI{3.75}{\micro\metre} with a total 1292 x 964 pixels. Our experimental setup shows 0.3\% noise that includes back-reflections by the equipment, stray light, the light source noise (0.2\%), the CCD camera quantification instability, and variation of experiment conditions. In our iterative optimization algorithm, we have targeted the beam of light between 420 nm - 535 nm (blue band) to the right spot and the light between 560 nm - 875 nm (red band) to the left spot at the diffraction plane.

\subsection{Generation of data set}

While developing a neural network for experimentally spectral splitting and spatially concentrating the broadband light, we generated a data set using the setup in Fig. \ref{experimental}a. The data collection procedure is followed as: first, we start with all SLM pixels having 0 phase shift. The SLM pixels are grouped by 30 x 30 forming a superpixel, and totally 64 x 36 superpixels exist. The phase of a superpixel is scanned from 0 to 2.28$\pi$ with a 0.23$\pi$ phase step. In the mean-time, we capture intensity distribution using a color camera. Later, we alter phase shift values of all SLM pixels sequentially and then collect intensity distributions by a color camera for each phase value of each SLM pixel. In order to perform iterative optimization, we write the phase value on the SLM that gives the highest intensity summation on red and blue target pixels. With this experimental configuration, we obtained 33796 phase plates (Fig. \ref{experimental}b) and corresponding intensity distributions (Fig. \ref{experimental}c) formed on the CCD camera. The experimental data set is collected within 1.7 hours that is used for training the neural network only once. After the training neural network generation of a phase plate for an intensity distribution of interest reduces to 2 seconds via the neural network. Considering the modeling duration of a neural network, we reduced the SLM pixels size by resulting in 36-by-64 total number of pixels. In a similar manner, we reduced the size of intensity distributions to a matrix size of 36-by-64. Here, we use percentage differential change (PDC) as a metric to indicate percentage increase in intensity at the target plane. $PDC^{x}(\lambda)$ at a pixel position of x for color band wavelength $\lambda$ is calculated via Eq. 1. $I_{i}^{x}(\lambda)$ and $I_{f}^{x}(\lambda)$ are the initial and final intensities at the pixel position of x and color band wavelength $\lambda$, respectively.

\begin{equation} \label{eq1}
PDC^{x}(\lambda) = 100*\frac{I_{f}^{x}(\lambda) - I_{i}^{x}(\lambda)}{I_{i}^{x}(\lambda)}
\end{equation}

\subsection{Neural network model}

\begin{figure}[!htb]
  \centering
    \includegraphics[width=85 mm]{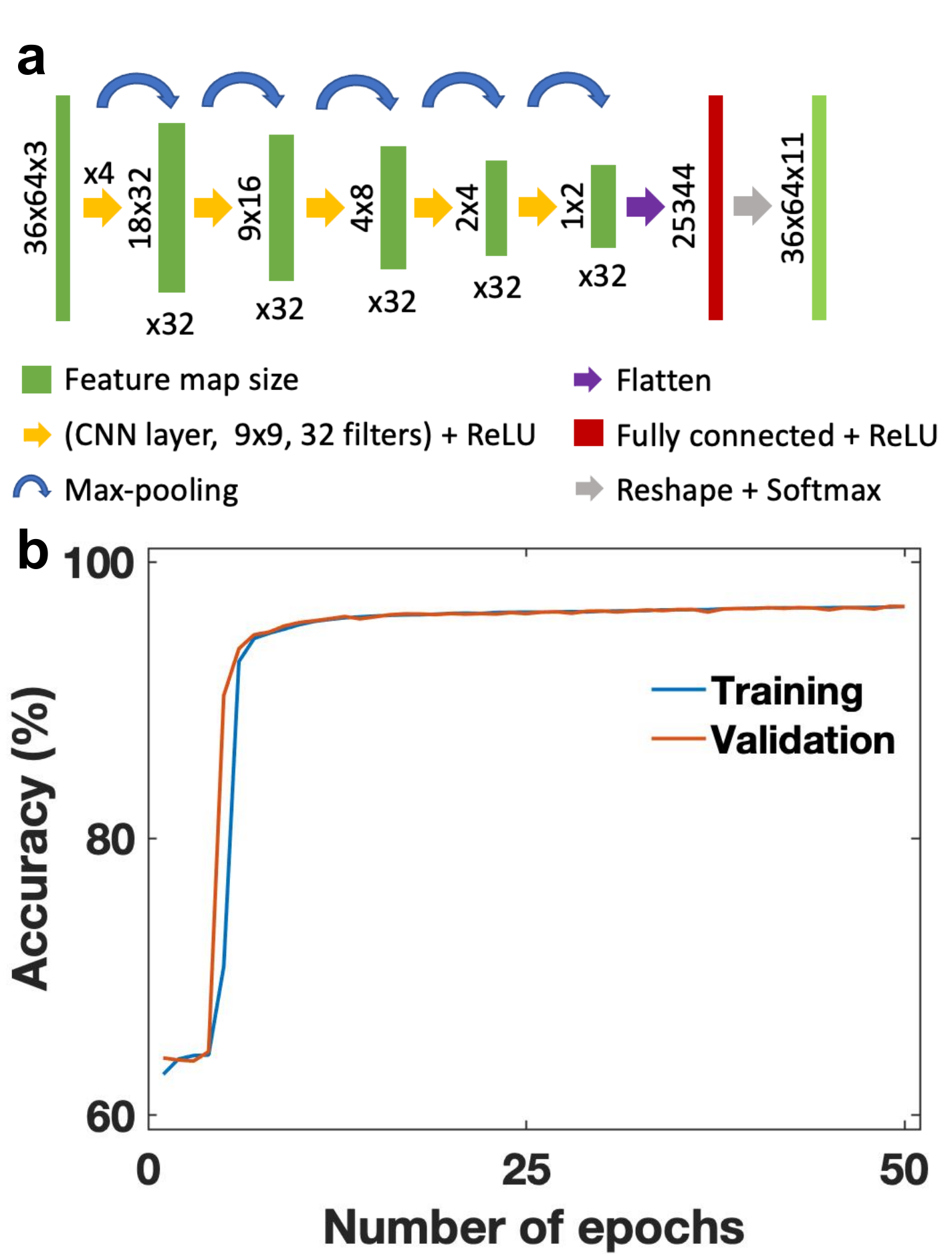}
    
  \caption{(a) Neural network model that is trained for experimentally spectral splitting and spatially concentrating the broadband light; (b) Training and validation accuracies of the neural network model with experimental data as a function of epochs, indicating that our model does not result in overfitting. }
  \label{modelaccuracy}
\end{figure}

Diffraction of light is expressed by Fresnel-Kirchhoff diffraction integral which makes it a suitable problem that can be addressed using convolutional neural network (CNN) layers. The relation between one pixel of the intensity distribution and one pixel of phase plate depends on many parameters. One pixel of the phase plates has a contribution to each pixel at the target plane \cite{Yolalmaz2020}. Moreover, when the input wavelength of the phase plate changes, the intensity distribution changes, and there is no explicit pattern between intensity on a target and the input wavelength. Thus, in our neural network model, we employed CNN layers to mimic relations between intensity distributions of formed diffraction patterns and corresponding phase patterns.

\begin{figure*}[!htb]
  
    
    \includegraphics[width=175 mm]{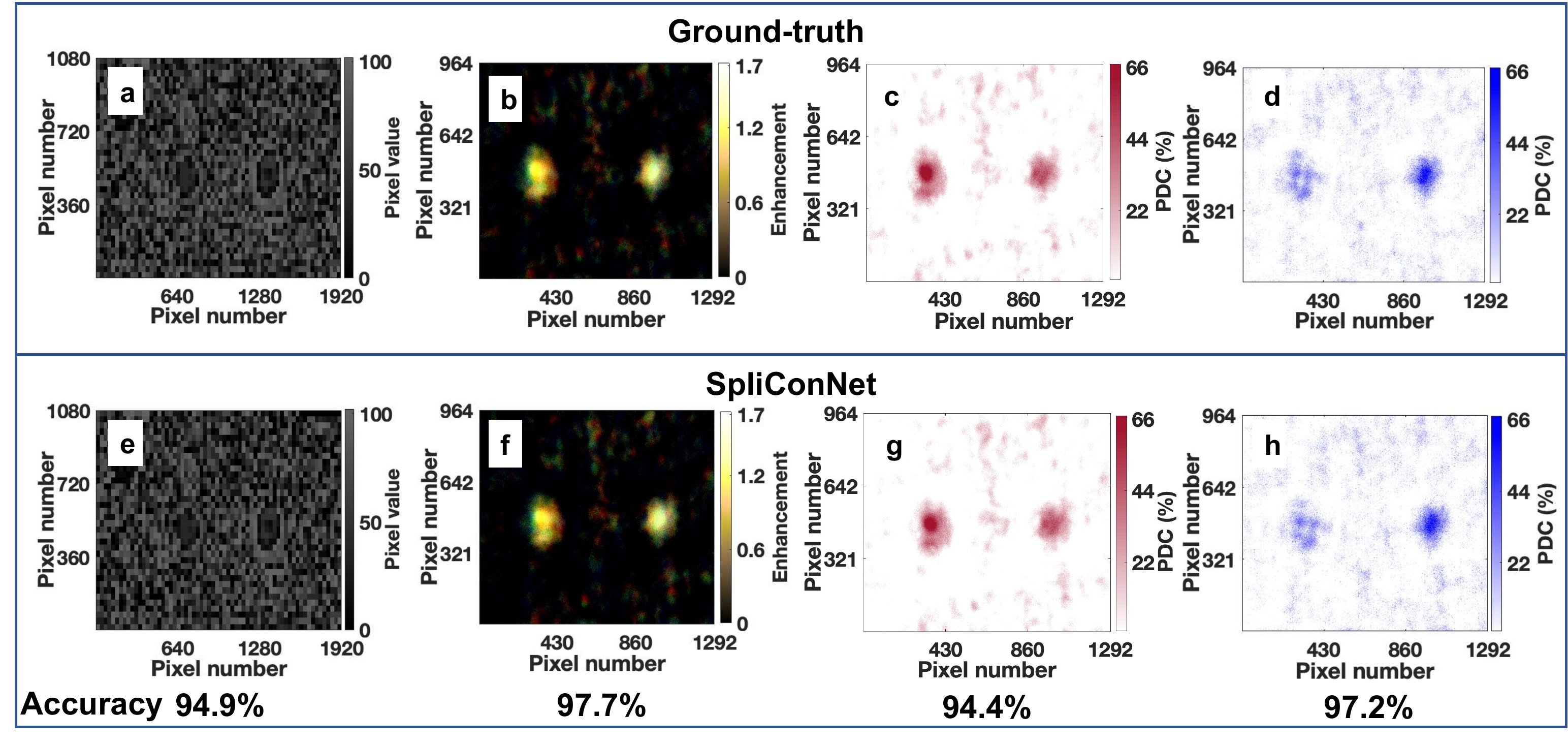}

  \caption{SpliCon that spectrally splits and concentrates two frequency bands; (a) Iteratively optimized phase pattern to split the broadband light into two bands on two regions; (b) Iteratively obtained intensity distribution of the broadband light on the colorful CCD camera; (c-d) the intensity distributions of the broadband light for red channel (between 560 nm - 875 nm) and blue channel (between 420 nm - 535 nm); (e) SpliConNet-based phase pattern; (f) SpliConNet-based intensity distribution of the broadband light on the CCD camera; (g-h) SpliConNet-based intensity distributions of the broadband light channels. Colors of the figures indicate color bands of the broadband light. PDC is percentage differential increase in intensity described in Eq. \ref{eq1}.}
  \label{results}
\end{figure*}

The neural network model developed for spectrally splitting and spatially concentrating the broadband light is presented in Fig. \ref{modelaccuracy}a. Using the aforementioned data generation protocol with the experimental setup, we fine-tuned the hyper-parameters and meta-parameters of this model. This model includes 8 CNN layers with a filter size of 9-by-9 and a filter number of 32. The CNN layers in the model are same-padded for keeping the size of feature maps invariant. After each CNN layer, an activation function of rectified linear unit (ReLU) is presented to reveal nonlinear relations between the intensity distributions and the phase plates. A ReLU activation function has an output of 0 if the input is less than 0; otherwise, ReLU activation function gives a raw output. After the 4th CNN layer, down-sampling of the feature maps is performed with a max-pooling operation at each CNN layer. After each max-pooling operation, the number of parameters and computation load in the network is reduced. A fully connected layer with a size of 25344 and a ReLU activation function is used after flattening operation. Then, we reshaped generated feature maps of the intensity distributions to match the size of the phase patterns. For the classification of the phase values of the phase patterns, we used a softmax activation function. It is a more generalized logistic activation function used in the output layer of a neural network for multi-label classification. Our batch size is selected as 32 for the smooth optimization of model weights. We call the network that we develop as Spectral Splitter and Concentrator Network (SpliConNet).

Using the versatile setup that we construct for training and testing SpliCons we collect 33796 camera images for training SpliConNet framework and use normalized intensity distributions in our model. We call Keras and Tensorflow open-source libraries, which provide tools of artificial neural networks in addition to GPU computing operation. We used the ADAM optimizer in Tensorflow to minimize categorical cross-entropy loss function over the training samples. Training of the model is completed in less than an hour using the Tensorflow library on a NVIDIA Quadro P5000 GPU. The latency of each training epoch is around 5 seconds. Once the training is completed, we test our model with validation data set which is 10\% of the input data set that is not part of the training set. The validation set prevents overfitting of the network model to the training set (Fig. \ref{modelaccuracy}b). After training, it takes only a few seconds to generate a phase plate for the desired intensity distribution.

\begin{figure}[!htb]
  
    
    \includegraphics[width=90 mm]{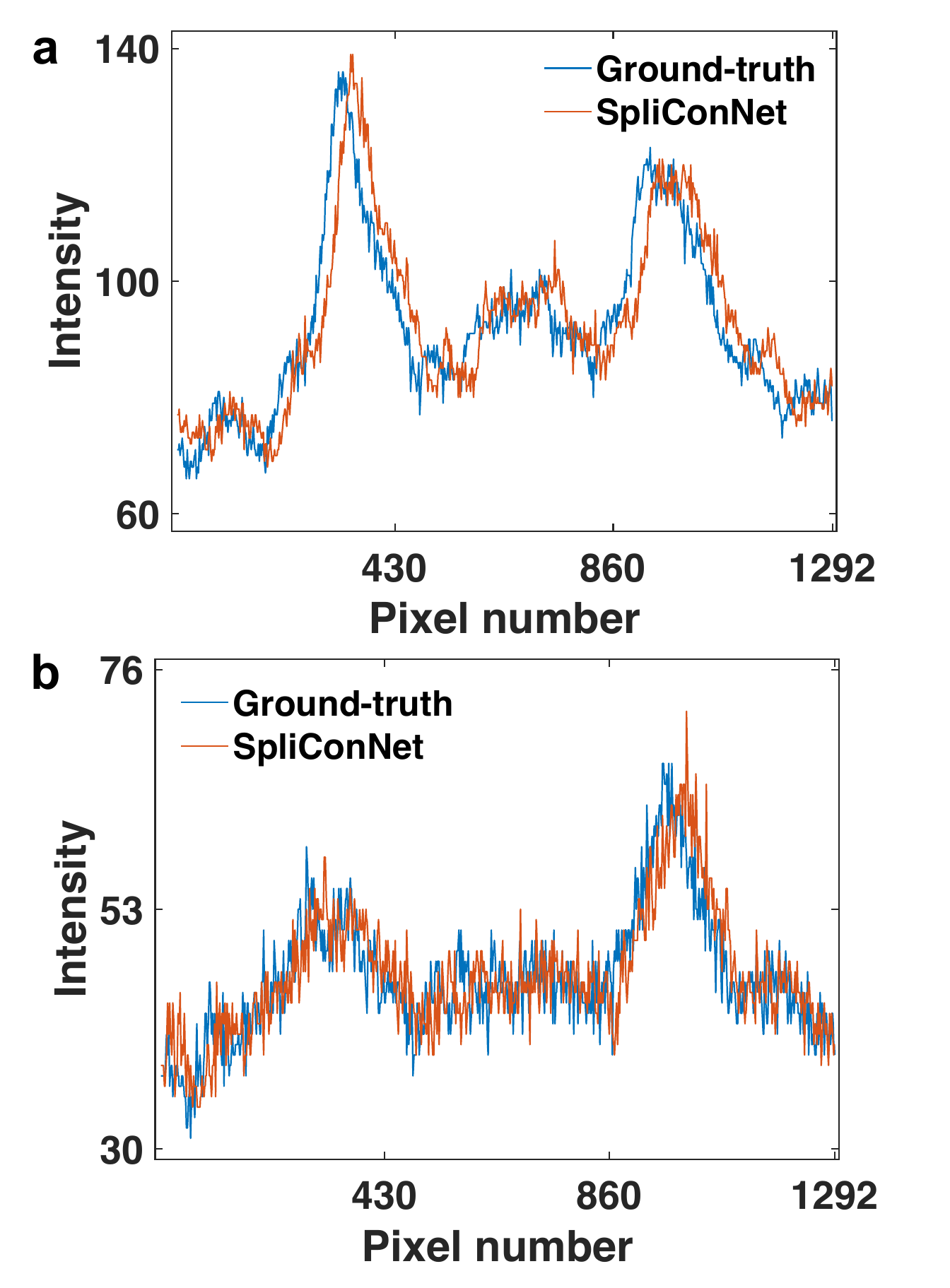}

  \caption{The output intensity patterns of the SpliCons for (a) red and (b) blue frequency bands that are optimized iteratively (Ground-truth) and via SpliConNet. The cross-sectional views are obtained at y=480 pixel along where color bands are concentrated.}
  \label{crosssection}
\end{figure}

\section{Results and discussion}

We carried out comprehensive experiments in order to both concentrate and spectrally split the broadband light using a SpliCon and the experimental setup seen in Fig. \ref{experimental}a. With the setup, we scan pixels of a phase plate to concentrate red and blue bands of the broadband light source on two targets. The phase plate that allows us to disperse the broadband light is presented in Fig. \ref{results}a, and it is the ground-truth phase plate. Using this ground-truth phase plate, we obtained intensity distribution of the light source as seen in Fig. \ref{results}b. This image is our ground-truth image to test the neural network model. This procedure is performed only once and takes about 2 hours. In Fig. \ref{results}b we provide the intensity distribution of the broadband light that is splitted and concentrated into two separate regions. The red band is concentrated on the left of the target plane (Fig. \ref{results}c) and the blue band is concentrated on the right of the target plane (Fig. \ref{results}d). 

Our goal is to use a neural network model to determine a function of the phase values of the phase pattern, $\phi$=f(I,w), where I is the intensity of wavefront shaped light, and f is a neural network model parametrized by a set of weights w. With the created data during the optimization of a phase plate, we trained the model in Fig. \ref{modelaccuracy}a. The model includes CNN layers to express function f in terms of weight w. Accuracies of the training and validation through epochs reach around 96.6 $\pm$ 2.3\% (Fig. \ref{modelaccuracy}b). With the results of this figure, we concluded that the weights of the neural network model are well-optimized, and the model lacks over-fitting as we reached similar accuracies with training and validation data sets. We test the performance of our neural network for splitting and concentrating the broadband light. When we reconstruct a phase pattern for the ground-truth CCD image (Fig. \ref{results}b) by using weights of the neural network, we obtain a similar phase pattern (Fig. \ref{results}e) with the phase pattern obtained by the experimental study (Fig. \ref{results}a). The agreement between these phase plates reaches up to 94.9\%, and this value depends on the accuracy of the model which is affected by the initialization of weights for the model. We saw 97.7 $\pm$ 2.7\% mean correlation between the reconstructed phase plates with all intensity distributions in the data set and the ground-truth phase plates.

With the reconstructed phase plate, we obtained a high correlation between the ground-truth CCD image (Fig. \ref{results}b) and the neural network-based CCD image (Fig. \ref{results}f) reaching up to 97.7 $\pm$ 0.3\% accuracy. The result that we obtain is limited by the setup noise of 0.3\%. Therefore, the method that we develop succeeds to reach the ground truth with unprecedented accuracy. With this neural network model we obtained excess 62.2\%  enhancement in red light band (Fig. \ref{results}g) and 61.0\% enhancement in blue light band (Fig. \ref{results}h) on the targets, respectively. We observe less than 4.8\% error in enhancement values of the light bands with the neural network compared to the experimental results. Considering the experimental setup noise of 0.3\%, the error we obtained in the CCD images is well in the expected regime.

In Fig. \ref{crosssection} we present variation of the intensities for two distinct frequency bands. Ground-truth results refer to color bands intensities of iteratively obtained CCD image. SpliConNet in the same figure corresponds to color bands intensities of the CCD image attained via SpliConNet developed. As can be seen in Fig. \ref{crosssection} we observe excellent agreement between the ground truth and the SpliConNet optimized intensity patterns.

Neural networks can better understand the fundamental science and drive knowledge discovery in addition to generating useful scientific output using comprehensive data sets. Identifying the input variables that are relevant for estimating the underlying function can assist researchers in better understanding the output of the problem. However, this may not provide information about the underlying physics. We think that physics-informed neural networks can be more beneficial in understanding the underlying physics \cite{Wiecha2021, Raissi2019}.  

Optimizing a phase plate for broadband light is quite a time-consuming process. When number of optimization parameters as the number of operating wavelength, number of pixels, etc. increase computation load is getting devastating. The number of parameters that we can control here reaches up to \num{2.5e4}, and experimental optimization using iterative methods takes up to 2 hours. The calculation of a broadband phase plate lasts approximately 89 days on a desktop PC, which is computationally unaffordable \cite{Yolalmaz2020}. Our neural network models infer phase patterns from intensity distributions obtained by Fresnel-Kirchhoff integral without any need for a prior mathematical model of the diffraction within a few seconds. The current approach presented in this manuscript is embodied by using a data-driven approach, neural network architecture. Our neural network reveals hidden information between the input and the output data. Thus, spectral and spatial characteristic of broadband light does not affect phase plate reconstruction capability of our neural network architecture. However, spatial coherence plays a crucial role in shaping the wavefront when a broadband light is used in measurements. The spatial coherence of the sunlight will provide the means to employ our method experimentally\cite{Divitt2015}. 

With the transfer learning tool we can significantly speed up the training procedure of our neural network model when new data set is fed from different setup schemes to reconstruct phase patterns for desired intensity distributions. Besides, we can inverse-design phase plates using our neural networks when the size of intensity distribution is up-scaled or down-scaled. Another important feature of our neural network-based spatial light concentration is to control spot size of modulated light. Further iterative optimization of these phase plates designs yields enhanced efficiencies, and we called this a hybrid technique that constitutes the local search optimization algorithm and the neural network model to improve reconstructed phase patterns of phase plates.

\section{Conclusions}

In this paper, we presented the design of SpliCons using a neural network model. Our model shows high accuracy to reconstruct phase patterns for spectrally splitting and spatially concentrating the broadband light. We obtained 97.7\% accuracy in CCD images and 94.9\% accuracy in phase plates. Using a reconstructed phase plate, we concentrate more than an excess of 61.0\% light on a target. We believe that the spectral and spatial control that we achieve here will pave the way for advanced applications in holography, microscopy, and information technologies in addition to solar energy harvesting. We openly share the fast and accurate framework that we develop in order to contribute to the design and implementation of difractive optical elements that will lead to transformative effects in diverse fields that require spatial and spectral control of light.

\section*{Supplementary Material}
See supplementary material for the SpliConNet architecture and data set.

\begin{acknowledgments}

This study is financially supported by The Scientific and Technological Research Council of Turkey (TUBITAK), grant no 118F075. PhD study of Alim Yolalmaz is supported by TUBITAK, with grant program of 2211-A. We would like to thank A. Tarık Temur and M. Ekrem Odabaş for discussions that improve our SpliConNet framework. We would like to thank Alpan Bek and Allard P. Mosk for providing us the essential equipment. We also would like to thank Raşit Turan and Ahmet Oral for their support during the establishment of our laboratory.

\end{acknowledgments}

\section*{DATA AVAILABILITY}
The data as well as the SpliConNet framework is available as a supplemental material.

\bibliography{Text}

\end{document}